\documentclass[reprint,amsmath,amssymb,aps,superscriptaddress,prd,nofootinbib, floatfix]{revtex4-2}
\usepackage{geometry}
\usepackage{graphicx}
\usepackage{mathtools}
\usepackage{siunitx}
\usepackage{xcolor}
\usepackage[normalem]{ulem}
\usepackage{url}
\usepackage{xr-hyper}
\usepackage{soul}
\usepackage{booktabs}
\usepackage{multirow}
\usepackage{subfiles}
\DeclareMathOperator{\sinc}{sinc}
\usepackage{comment}
\usepackage{lineno}
\begin{document}

\title{Beyond the Long Wavelength Approximation: Next-generation Gravitational-Wave Detectors and Frequency-dependent Antenna Patterns}

\author{Andrea Virtuoso}
\email[Correspondence email address: ]{andrea.virtuoso@ts.infn.it}
\affiliation{INFN Sezione di Trieste, I-34127 Trieste, Italy} 
\author{Edoardo Milotti}
\affiliation{INFN Sezione di Trieste, I-34127 Trieste, Italy} 
\affiliation{Università di Trieste, Dipartimento di Fisica, I-34127 Trieste, Italy}

\date{\today}

%-----------------------------
%------------ Abstract -------------
%-----------------------------
\begin{abstract}

The response of a gravitational--wave (GW) interferometer is spatially modulated and is described by two antenna patterns, one for each polarization state of the waves. The antenna patterns are derived from the shape and size of the interferometer, usually under the assumption that the interferometer size is much smaller than the wavelength of the gravitational waves (long wavelength approximation, LWA). This assumption is well justified as long as the frequency of the gravitational waves is well below the free spectral range (FSR) of the Fabry-Perot cavities in the interferometer arms as it happens for current interferometers ($\mathrm{FSR}=37.5$~kHz for the LIGO interferometers and $\mathrm{FSR}=50$~kHz for Virgo and KAGRA). However, the LWA can no longer be taken for granted with third--generation instruments (Einstein Telescope, Cosmic Explorer and LISA) because of their longer arms. This has been known for some time, and previous analyses have mostly been carried out in the frequency domain. In this paper, we explore the behavior of the frequency--dependent antenna patterns in the time domain and in the time--frequency domain, with specific reference to the searches of short GW transients. We analyze the profound changes in the concept of Dominant Polarization Frame, which must be generalized in a nontrivial way, we show that the conventional likelihood-based analysis of coherence in different interferometers can no longer be applied as in current analysis pipelines, and that methods based on the null stream in triangular ($\ang{60}$) interferometers no longer work. Overall, this paper establishes methods and tools that can be used to overcome these difficulties in the unmodeled analysis of short GW transients. 
\end{abstract}

\maketitle

\section{Introduction}\label{sec:introduction}

After the discovery of the first gravitational wave GW150914 \cite{abbott2016observation,abbott2016observing, abbott2016gw150914}, the LIGO-Virgo-KAGRA (LVK) collaboration has detected and analysed an ever increasing number of gravitational wave (GW) events \cite{abbott2019gwtc1, abbott2021gwtc2, abbott2021gwtc3}.
This effort is leading to a deeper and deeper understanding of the Universe, and GWs are not just an object of study by themselves, but have become new powerful tools to infer the properties of GW source populations \cite{abbott2023population}, test the limits of the Theory of General Relativity (GR) \cite{abbott2021tests}, and much more.

To do this, the LVK collaboration processes the strain data from the detectors with several analysis pipelines, which can be divided into two main classes: \textit{modeled methods}, which use theoretical waveforms and matched filters to detect and analyse GWTs \cite{veitch2015parameter,usman2016pycbc,dal2021real,aubin2021mbta,tsukada2023improved}, and \textit{unmodeled} (or \textit{minimally modeled}) \textit{methods}, which are  based instead on the coherence of data in different detectors and do not assume a specific signal morphology \cite{sutton2010x,klimenko2016method,drago2021coherent, millhouse2018bayesian,cornish2021bayeswave}.

While these pipelines can be very different from one other, they all model the detector response as follows
\begin{equation}\label{eq:detector_response_first_formula}
    \xi_{F}(t,\hat{\textbf{n}})=F_{+}(\hat{\textbf{n}})h_{+}(t)+F_{\times}(\hat{\textbf{n}})h_{\times}(t),
\end{equation}
where $h_{+,\times}$ are the two polarizations predicted by the Theory of General Relativity \cite{abbott2021tests} the coefficients $F_{+,\times}$ are the antenna patterns, which depend on the source location $\hat{\textbf{n}}$ as seen in each detector frame
\iffalse
\begin{align}
    F_{+}(\hat{\textbf{n}})&=\dfrac{1}{2}\left(e_{11}^{+}(\hat{\textbf{n}})-e_{22}^{+}(\hat{\textbf{n}})\right)
    F_{\times}(\hat{\textbf{n}})&=\dfrac{1}{2}\left(e_{11}^{\times}(\hat{\textbf{n}})-e_{22}^{\times}(\hat{\textbf{n}})\right)=-\cos{\theta}\sin{2\phi},\label{eq:usual_antenna_pattern_cross}
\end{align}
\fi 
\begin{equation}
\label{eq:usual_antenna_pattern}
    F_A(\hat{\textbf{n}}) = \frac{1}{2} \epsilon_{ij}^A(\hat{\textbf{n}}) \left( e_x^i e_x^j - e_y^i e_y^j \right) \quad (A=+,\times),
\end{equation} 
where $\epsilon^{A}_{ij}(\hat{\textbf{n}})$ are the polarization tensors and $e_x^i$, $e_y^i$ are the unit vectors that define the interferometer arm directions.
Eq. \eqref{eq:detector_response_first_formula}, and the expression for the antenna patterns \eqref{eq:usual_antenna_pattern} 
%and \eqref{eq:usual_antenna_pattern_cross}, 
are based on the assumption that the  wavelength of the gravitational wave is much larger than the length of the detector arms \cite{schutz2009first} -- the so-called \emph{long wavelength approximation} (LWA). 
As a result, the antenna patterns \eqref{eq:usual_antenna_pattern} are frequency--independent. 
The physical interpretation of the LWA is that the signal is nearly constant during a photon round trip in each interferometer arm or equivalently that its frequency is well below the detector free spectral range, defined as $f_\mathrm{FSR}=c/2L$, where $L$ is the length of the Fabry-Perot cavities in the interferometer arms.
This hypothesis is well-justified for the current detectors. Indeed, taking the LIGO Livingston detector as a standard, we find that in the third observing run (O3) the strain sensitivity is better than $10^{-23}/\sqrt{\mathrm{Hz}}$ in the range $\sim \SI{50}{Hz}$-$\SI{1000}{Hz}$, values which are well below LIGO's free spectral range $f_\mathrm{FSR}=\SI{37500}{Hz}$ (i.e., $f/f_\mathrm{FSR} \lesssim 0.027$).

This shall not be the case for the next generation of ground-based detectors, such as Einstein Telescope (ET) \cite{maggiore2020science, branchesi2023science} and Cosmic Explorer (CE) \cite{reitze2019cosmic, evans2021horizon, hall2022cosmic}, both designed to considerably improve the sensitivity of current detectors, opening a wide range of opportunities to understand our Universe with GWs.
The sensitivity goal will be achieved also by building arms significantly longer than those of current detectors, although the exact length is not known at the time of writing, since the design of both ET and CE (as well as the number of detectors) is not yet frozen, and many options are currently under discussion.
In short, the decision for ET is between the following alternatives \cite{branchesi2023science}
\begin{itemize}
    \item A triangular design consisting of three double interferometers in the same location having arms with a length of $\SI{10}{km}$ (or possibly $\SI{15}{km}$), with an angle of $\ang{60}$ between the arms.
    \item An L design consisting of two well-spaced interferometers, each with orthogonal arms of length $\SI{15}{km}$ (or possibly $\SI{20}{km}$); the two detectors should be oriented $\sim \ang{45}$ with respect to each other (or possibly with the arms of the two interferometers forming the same angle with respect to the great circle connecting the two detectors).
\end{itemize}
Both configurations are designed to operate over a wide range of frequencies: from a few $\SI{}{Hz}$ up to thousands $\SI{}{Hz}$.

CE \cite{evans2021horizon} should have an L design with arms at $\ang{90}$: the possible alternatives are two $\SI{40}{km} $ detectors or a pair with a $\SI{40}{km}$ detector and a $\SI{20}{km}$ detector: the latter configuration would be tuned to be more sensitive in the high frequency range (above $\SI {1000} {Hz}$) where we expect to observe the post-merger phase of BNS.

\medskip

With the increase in arm length, as well as with the increase in sensitivity to high frequencies due to other technological advances, we need to reconsider the validity of the LWA.
Although we delve into this topic in much more detail below, we start by examining a few simple figures. 
Consider the ET configuration with the shortest planned arm length, i.e., the triangular design with $L=\SI{10}{km}$: its free spectral range is $f_\mathrm{FSR}=\SI{15000}{Hz}$ and therefore a signal frequency $f=\SI{2000}{Hz}$ implies the ratio $f/f_\mathrm{FSR}\simeq 0.13$, which is no longer negligible. This ratio is more impacting on the CE configurations, where a signal frequency $f=\SI{1000}{Hz}$ implies the ratios $f/f_\mathrm{FSR} \simeq 0.13$ for the $\SI{20}{km}$ arms and $f/f_\mathrm{FSR} \simeq 0.27$ for the $\SI{40}{km}$ arms.

\medskip 

These non negligible ratios indicate that with the new detectors we must abandon the LWA. This is already known, and the issue has been discussed in depth in the frequency domain since the early 2000s \cite{rakhmanov2008high,LIGODocumentT070172-x0,LIGODocumentT060237,LIGODocumentT080134,LIGODocumentT080134} as well as, among others, in a recent study dedicated to the next generation of detectors \cite{essick2017frequency}.
These studies confirm and extend the considerations on the frequency dependence of the antenna patterns introduced above.
The LISA scientific community is aware of the problem  \cite{cornish2003lisa, kudoh2005probing, babak2021lisa}, because the planned arm length is huge ($\SI{2.5}{Mkm}$, see \cite{baker2019laser}) and the LWA has been dropped since the initial LISA studies and simulations.

On the contrary, little or no attention has been given, even within the LISA scientific community, to the time-domain description of the detector response beyond the LWA, as well as to the combined time-frequency response.
While modeled analysis is often implemented in the frequency domain, both in pipelines analyzing data from current interferometers \cite{veitch2015parameter,usman2016pycbc,dal2021real,aubin2021mbta,tsukada2023improved} and in planned methods for future interferometers like LISA \cite{arnaud2006mock, babak2008mock, arnaud2007overview, babak2008report, babak2010mock}, the time domain and, above all, the time-frequency domain methods are at the basis of unmodeled analysis of GW transients \cite{abbott2017all, abbott2019all, abbott2021all, klimenko2016method,drago2021coherent,millhouse2018bayesian,cornish2021bayeswave, szczepanczyk2021observing, szczepanczyk2023search}.

Under general conditions, we show that both in the time domain and in the time--frequency domain it becomes impossible to disentangle the polarized strains from the antenna patterns. In current unmodeled pipelines, this disentangling is necessary to the analysis and can be performed thanks to the frequency--independent antenna patterns obtained assuming the LWA. Here, we obtain the conditions under which a traditional analysis with frequency--independent antenna patterns can still be performed with next--generation detectors.  We also reconsider in detail  concepts that are important in a likelihood--based unmodeled analysis, like the Dominant Polarization Frame and the null stream.

\medskip 

These considerations do not just impact unmodeled methods \textit{per se} but also the physical and astrophysical results of GW signal analysis. For instance, they are relevant in all applications of the unmodeled methods to tests of General Relativity \cite{abbott2021tests}, since their neglect can lead to significant distortions of the reconstructed waveforms. Moreover, while compact binary coalescences can be studied with great precision with modeled methods based on matched filters thanks to the exceptional accuracy of the theoretical results, other astrophysical phenomena yet undetected in the GW domain, like core--collapse supernovae, produce GW signals which are not completely deterministic \cite{kotake2009stochastic,powell2022inferring,szczepanczyk2021detecting,szczepanczyk2023optically} and we expect unmodeled detections and analyses to be of pivotal importance in such cases. Next--generation detectors are also expected to detect GW signals from the post--merger of BNS coalescences, which are characterized by frequencies for which the LWA would not hold \cite{topolski2023post}.

\bigskip

In the following sections we describe the detector response beyond the LWA both in the time domain and in the time--frequency domain, we compare them with the frequency--domain formula, and we discuss their interpretation and applicability to gain a comprehensive view of their implications.

Next, we consider the impact of the frequency-dependent antenna patterns on a specific embodiment of unmodeled analysis based on likelihood, as in the case of \textit{coherent WaveBurst} (cWB), an important pipeline used by the LVK Collaboration \cite{klimenko2016method,drago2021coherent}. 
We do so by comparing the original formalism using frequency-independent antenna patterns, as in cWB, with an adjusted formalism based on frequency-dependent antenna patterns.
In this context, we show that the use of frequency-independent antenna patterns to reconstruct a gravitational waveform can lead to distortions that depend both on the frequency range of the gravitational-wave signal and on the sky location of the source. These distortions shall become important in next-generation detectors and disappear if we replace the frequency-independent antenna patterns with their frequency-dependent versions.

Finally, we discuss the null stream \cite{gursel1989near} in next--generation detectors. This is particularly relevant for the triangular design of ET, where it is expected to obtain a null stream which is independent of the sky location of the source \cite{wen2005coherent}.
While this is true in the LWA, we find that this is no longer the case when we move beyond the LWA and consider frequency-dependent antenna patterns.

\section{The detector response beyond the  LWA}\label{sec:detector_response}

\subsection{The frequency domain formula}

For the sake of completeness, we recall the well-known frequency-domain formula for the response of a gravitational-wave detector based on a Michelson interferometer with $90^\circ$ arms, with Fabry-Perot cavities in its arms \cite{rakhmanov2008high,LIGODocumentT070172-x0,LIGODocumentT060237,LIGODocumentT080134,LIGODocumentT080134, essick2017frequency}.
We use the convention of Sathyaprakash and Schutz \cite{sathyaprakash2009physics} for the direction to the source $\hat{\textbf{n}}$, and we denote $\hat{\textbf{x}}$ and $\hat{\textbf{y}}$ as the unit vectors associated with the two arms.
Denoting the detector arm length with $L$, the time of flight of photons in the arm is $T=L/c$, and we define    $T^{\hat{\textbf{a}}}_{\pm}=T(1\pm\hat{\textbf{a}}\cdot\hat{\textbf{n}})$, with $\hat{\textbf{a}}=\hat{\textbf{x}}, \hat{\textbf{y}}$\footnote{It is easy to see that $T^{\hat{\textbf{a}}}_{+}+T^{\hat{\textbf{a}}}_{-}=2T$.}.
Then, the detector response can be written as\footnote{We use the symbol $G$ to denote the  frequency-dependent antenna patterns, and label the detector response with the subscripts $F$ and $G$, respectively in the frequency--independent and in the frequency--dependent case. Moreover, the tilde is used to denote Fourier transforms and by extension all the frequency--dependent functions.}
\begin{equation}\label{eq:frequency_dependent_antenna_patterns_frequency_detector_response}
    \tilde{\xi}_{G}(f,\hat{\textbf{n}}) = \sum_{A=+,\times} \tilde{G}_{A}(f,\hat{\textbf{n}}) \tilde{h}_{A}(f),
\end{equation}
where 
\begin{align}
    \tilde{G}_{+}(f,\hat{\textbf{n}})& \equiv \frac{1}{2} \left[ \tilde{D}(\hat{\textbf{x}},\hat{\textbf{n}},f)\epsilon_{11}^{+}(\hat{\textbf{n}})
    -\tilde{D}(\hat{\textbf{y}},\hat{\textbf{n}},f) \epsilon_{22}^{+}(\hat{\textbf{n}}) \right]\label{eq:frequency_dependent_antenna_pattern_plus_definition}\\
    \tilde{G}_{\times}(f,\hat{\textbf{n}})& \equiv \frac{1}{2} \left[ \tilde{D}(\hat{\textbf{x}},\hat{\textbf{n}},f)\epsilon_{11}^{\times}(\hat{\textbf{n}})
    -\tilde{D}(\hat{\textbf{y}},\hat{\textbf{n}},f) \epsilon_{22}^{\times}(\hat{\textbf{n}}) \right]\label{eq:frequency_dependent_antenna_pattern_cross_definition}
\end{align}
 are the frequency-dependent antenna patterns, and where $\tilde{D}$ is the transfer function\footnote{This expression of the transfer function coincides with the one reported in \cite{rakhmanov2009round} except for the subscript of the $T^{\hat{\mathbf{a}}}$, which is opposite in \cite{rakhmanov2009round} because of the different convention on $\hat{\mathbf{n}}$.}
\begin{multline}\label{eq:D_transfer_function_definition}
     \tilde{D}(\hat{\textbf{a}},\hat{\textbf{n}},f) 
     =\frac{e^{-i2\pi f T}}{2}  \\
     \times  \left[
     e^{i\pi f T_{-}^{\hat{\textbf{a}}}}\sinc{(\pi f T_{+}^{\hat{\textbf{a}}})} 
     + e^{-i\pi f T_{+}^{\hat{\textbf{a}}}}
    \sinc{(\pi f T_{-}^{\hat{\textbf{a}}})}   
     \right].
\end{multline}  

The transfer function $\tilde{D}(\hat{\textbf{a}},\hat{\textbf{n}},f)$ is key to understanding the limitations of the LWA, as it depends on the arm length $L$, the signal frequency $f$ and on the sky location of the GW source $\hat{\textbf{n}}$. 
Here, we discuss its behaviour considering both current and next generation detectors.
For the latter, we take as a reference the designs presented in the Introduction. In particular, we denote as 
\begin{itemize}
    \item ET1\_L and ET2\_L the two $\ang{90}$ interferometers with arm length $L=\SI{15}{km}$ planned for the ET 2L design 
    \item ET1\_tr, ET2\_tr and ET3\_tr the three $\ang{60}$ interferometers with arm length $L=\SI{10}{km}$ planned for the ET triangular design
    \item CE1 and CE2 the two $\ang{90}$ interferometers with arm length, respectively, $L=\SI{40}{km}$ and $L=\SI{20}{km}$ planned for CE.
\end{itemize}

\begin{figure}[h!]
    \centering
    \includegraphics[width=0.95\columnwidth]{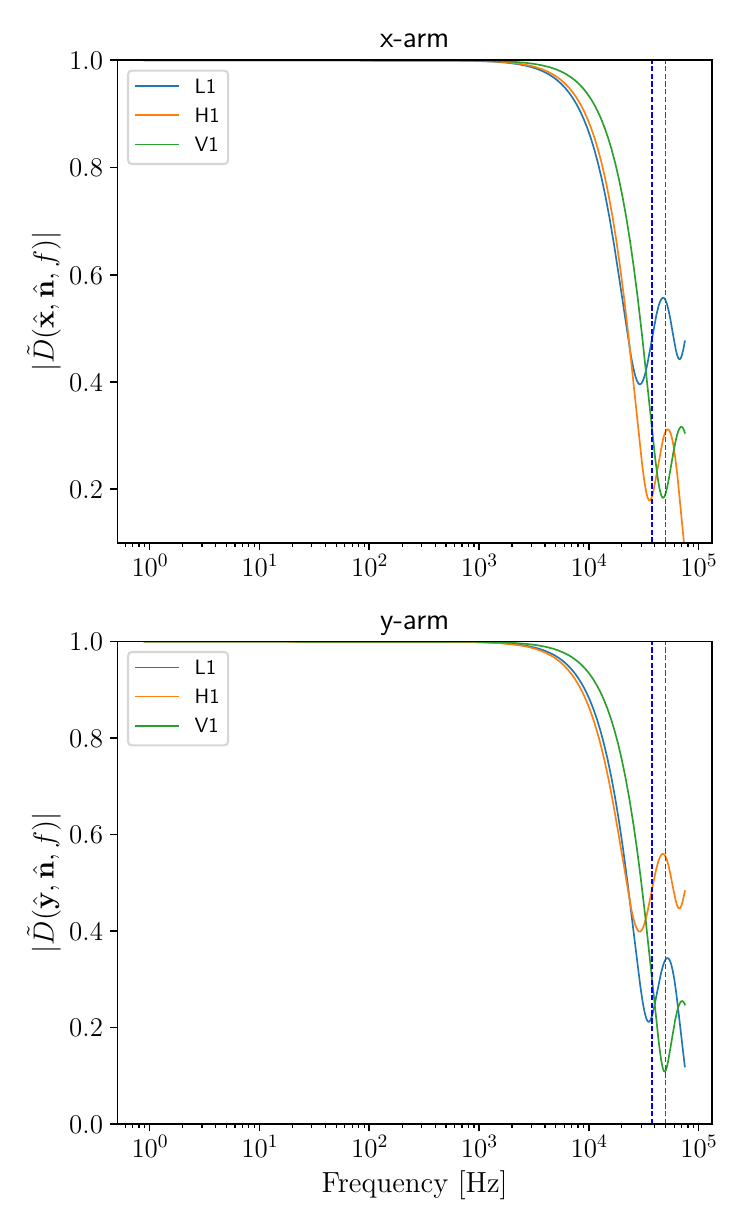}
    \caption{Absolute value of the transfer function $\tilde{D}(\hat{\textbf{a}},\hat{\textbf{n}},f)$ for LIGO and Virgo detectors at the source location $(\theta, \phi)=(\ang{60}, \ang{30})$, for both arms. The dashed vertical lines represent the FSR frequency (blue, LIGO detectors; green Virgo). As expected, the absolute value of the transfer function is nearly constant -- and very close to $1$ -- for all frequencies up to about $\SI{1000}{Hz}$, which shows that the LWA is well justified for all signals detected in the first three observing runs. }
    \label{fig:D_transfer_function_current_detectors}
\end{figure}

\begin{figure}[h!]
    \centering
    \includegraphics[width=0.95\columnwidth]{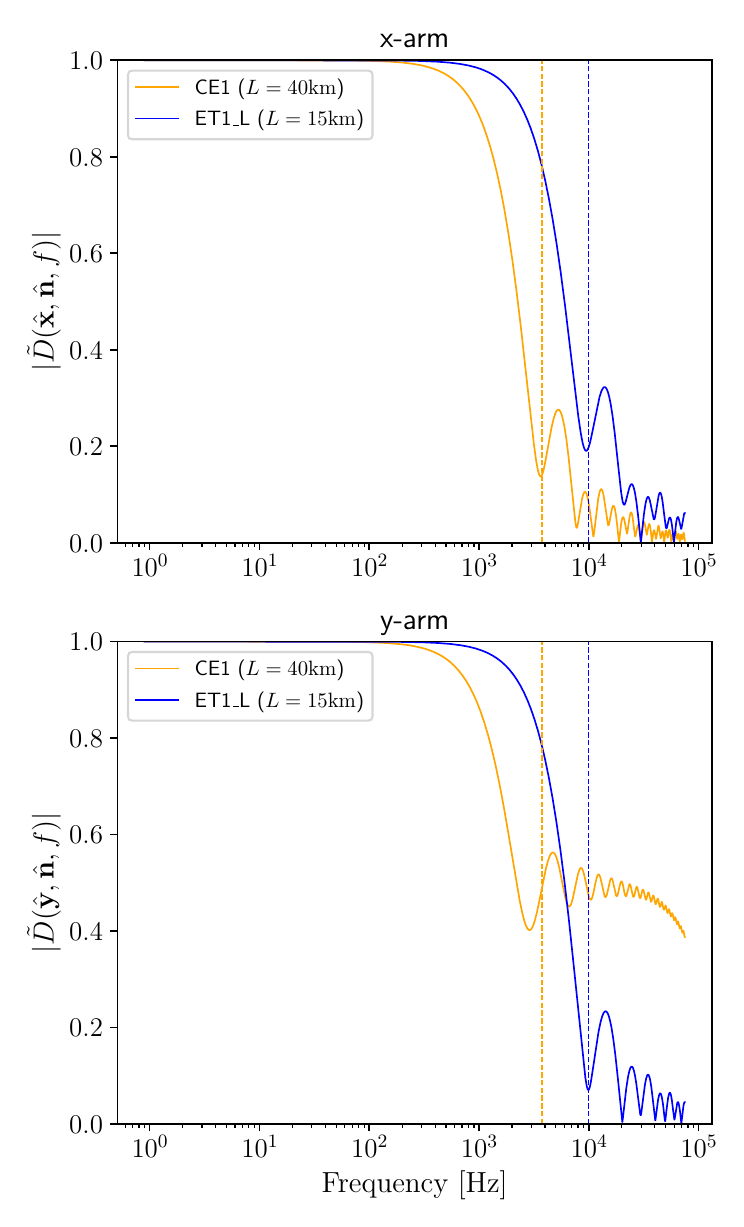}
    \caption{Absolute value of the transfer function $\tilde{D}(\hat{\textbf{a}},\hat{\textbf{n}},f)$ for CE1 (Idaho, $L=\SI{40}{km}$) and ET1 (Sardinia, $L=\SI{10}{km}$, L-shaped) at the source location $(\theta, \phi)=(\ang{60}, \ang{30})$, for both arms. The dashed vertical lines represent the FSR frequency (blue, ET1; orange CE1).
    Here, the absolute value of the transfer function is very different from that of current detectors (Figure \ref{fig:D_transfer_function_current_detectors}) because of the longer arm length, which causes the absolute value to drop from $1$ much earlier than $\SI{1000}{Hz}$.}
    \label{fig:D_transfer_function_next_gen_detectors}
\end{figure}

At the time of this writing, no decision as been taken yet on the location of the detectors, therefore we have selected a few of the sites currently under discussion \cite{evans2021horizon,branchesi2023science,borhanian2021gwbench}: we collect these and other details in Section G in the Supplementary Text.
The different detector designs lead to different transfer functions $\tilde{D}(\hat{\textbf{a}},\hat{\textbf{n}},f)$, and here we plot $|\tilde{D}(\hat{\textbf{a}},\hat{\textbf{n}},f)|$ for different detectors and for a specific source location $(\theta,\phi)=(\ang{60}, \ang{30})$, see Fig. \ref{fig:D_transfer_function_current_detectors} and Fig. \ref{fig:D_transfer_function_next_gen_detectors}. 
As expected, in current detectors the transfer function is very nearly constant up to about $\sim\SI{1000}{Hz}$ while in next-generation detectors the transfer function starts dropping at about $\sim\SI{200}{Hz}$ (see Figure \ref{fig:D_transfer_function_next_gen_detectors}).

\begin{figure}[ht!]
    \centering
    \includegraphics[width=0.9\textwidth]{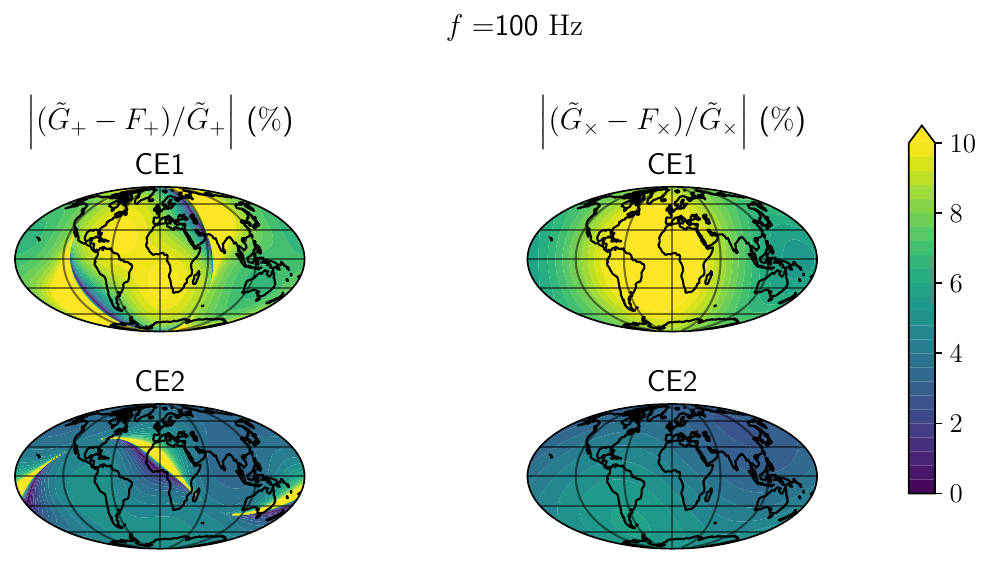}
    \caption{Fractional difference $|(\tilde{G}_{A}-F_{A})/\tilde{G}_{A}|$ ($A=+,\times$) evaluated as \% for $f=\SI{100}{Hz}$ for CE1 and CE2. This difference is mostly in the $0\%$--$10\%$ range for CE1, with  the corrections impacting only part of the sky.}
    \label{fig:percentage_difference_100_Hz_CE_detectors}
\end{figure}

\begin{figure}[ht!]
    \centering
    \includegraphics[width=0.9\textwidth]{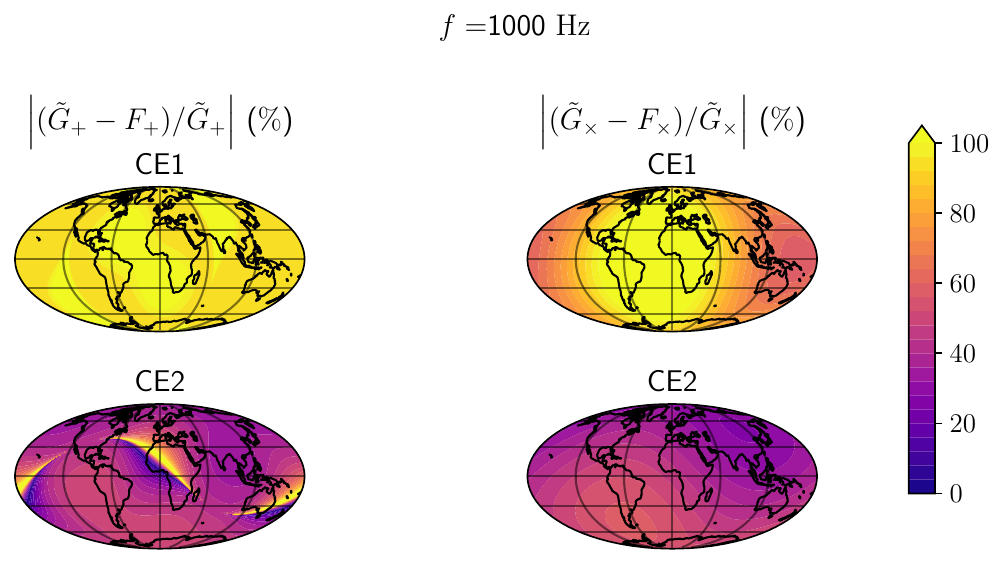}
    \caption{Fractional difference $|(\tilde{G}_{A}-F_{A})/\tilde{G}_{A}|$ ($A=+,\times$) evaluated as \% for $f=\SI{1000}{Hz}$ considering the next-generation detectors CE1, CE2. This difference is almost everywhere between $10\%$ and $100\%$ (and somewhere even higher for CE1 due to its considerably longer arms), therefore the high-frequency corrections are always needed to avoid important biases in data analysis.}  \label{fig:percentage_difference_1000_Hz_CE_detectors}
\end{figure}

\begin{figure}[h!]
    \centering
    \includegraphics[width=0.9\textwidth]{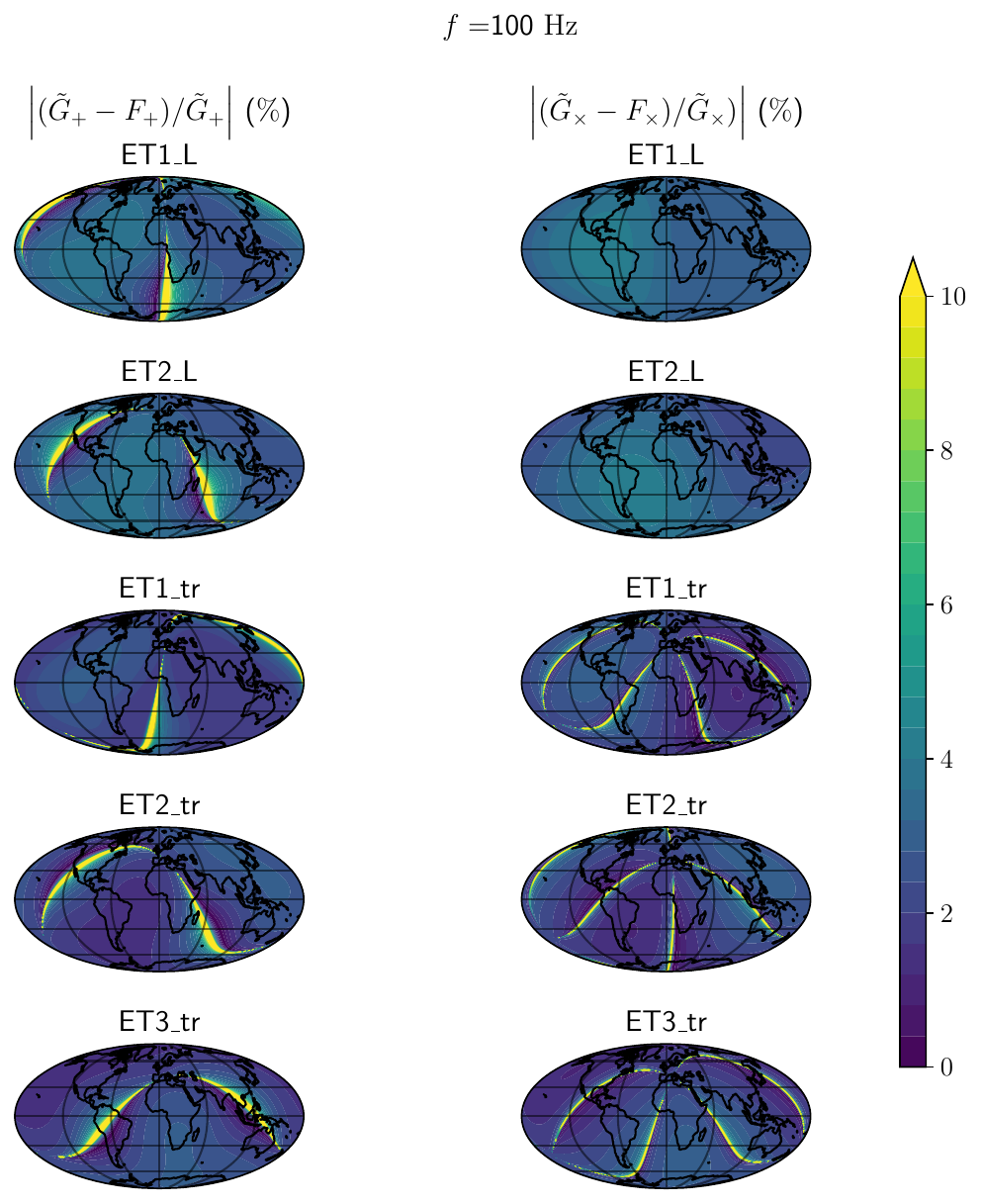}
    \caption{Fractional difference $|(\tilde{G}_{A}-F_{A})/\tilde{G}_{A}|$ ($A=+,\times$) evaluated as \% for $f=\SI{100}{Hz}$ for ET1\_L, ET2\_L, ET1\_tr, ET2\_tr and ET3\_tr. This difference is almost everywhere between $0\%$ and $10\%$, so that the high-frequency corrections are relevant only for some source locations.}
    \label{fig:percentage_difference_100_Hz_ET_detectors}
\end{figure}

\begin{figure}[h!]
    \centering
    \includegraphics[width=0.9\textwidth]{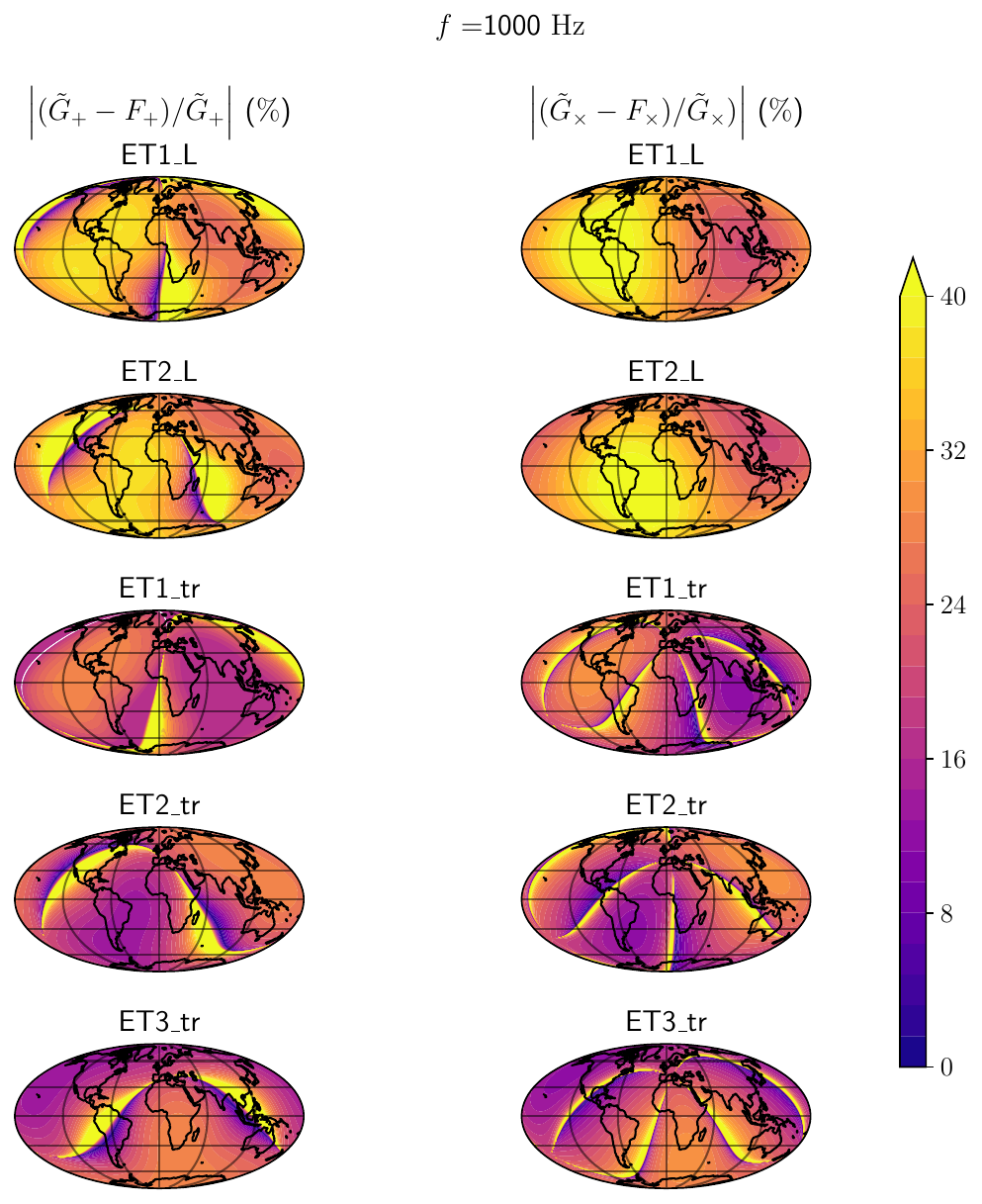}
    \caption{Fractional difference $|(\tilde{G}_{A}-F_{A})/\tilde{G}_{A}|$ ($A=+,\times$) evaluated as \% for $f=\SI{1000}{Hz}$ for ET1\_L, ET2\_L, ET1\_tr, ET2\_tr and ET3\_tr. This difference is almost everywhere between $10\%$ and $50\%$ (and often higher), therefore the high-frequency corrections are always needed to avoid important biases in data analysis.}  \label{fig:percentage_difference_1000_Hz_ET_detectors}
\end{figure}

These results hold for a specific sky location, but the frequency dependence changes as we move to a different sky location. Moreover, we also wish to evaluate the difference between the frequency-dependent antenna patterns of eqs. \eqref{eq:frequency_dependent_antenna_pattern_plus_definition} and \eqref{eq:frequency_dependent_antenna_pattern_cross_definition} and the conventional frequency-independent antenna patterns obtained assuming the LWA. 
{}
{}
To this end, we plot the fractional difference between frequency-dependent and frequency-independent antenna patterns -- i.e. $|(\tilde{G}_{A}-F_{A})/\tilde{G}_{A}|$ (for both $A=+,\times$) -- for $f=\SI{100}{Hz}$ and $f=\SI{1000}{Hz}$, for both CE (Figures \ref{fig:percentage_difference_100_Hz_CE_detectors} and \ref{fig:percentage_difference_1000_Hz_CE_detectors}) and ET (Figures \ref{fig:percentage_difference_100_Hz_ET_detectors} and \ref{fig:percentage_difference_1000_Hz_ET_detectors}).
The difference is not constant over the sky, meaning that there are some source locations for which the high-frequency corrections are more relevant.
For $f=\SI{100}{Hz}$, there are regions of the sky with a difference close to 10\%, especially for CE1; for $f=\SI{1000}{Hz}$, the difference is considerably higher for all detectors and for nearly all source locations.
This initial analysis suggests that for the next generation of the detectors at frequencies close to $f=\SI{1000}{Hz} $ or higher, the frequency-dependent corrections should be included in the analysis pipelines.

\newpage

\onecolumngrid

{}
{}

{}

{}

\twocolumngrid

\clearpage
\subsection{The time domain formula}

We are now ready to examine the time-domain version of the detector response beyond the LWA, which is given by the following formula:
\begin{widetext}  
\begin{align}\label{eq:frequency_dependent_antenna_patterns_time_detector_response}
    \xi_{G}(t,\hat{\textbf{n}})=\sum_{A=+,\times}&\dfrac{1}{2}\epsilon_{11}^{A}(\hat{\textbf{n}}) 
    \left(  \dfrac{1}{2T_{-}^{\hat{\textbf{x}}}}\int_{t-2T}^{t-T_{+}^{\hat{\textbf{x}}}} h_{A}(t')dt'
    +\dfrac{1}{2T_{+}^{\hat{\textbf{x}}}}\int_{t-T_{+}^{\hat{\textbf{x}}}}^{t} h_{A}(t')dt' \right) \nonumber \\
    -&\dfrac{1}{2}\epsilon_{22}^{A}(\hat{\textbf{n}})  
    \left(  \dfrac{1}{2T_{-}^{\hat{\textbf{y}}}}\int_{t-2T}^{t-T_{+}^{\hat{\textbf{y}}}} h_{A}(t')dt'
    +\dfrac{1}{2T_{+}^{\hat{\textbf{y}}}}\int_{t-T_{+}^{\hat{\textbf{y}}}}^{t} h_{A}(t')dt' \right).
\end{align} 
\end{widetext}
Its derivation and the meaning of all symbols are given in the Supplementary Text, Section A, while in Section C we show the equivalence of time-domain and frequency-domain representations.
A noteworthy feature of the general formula \eqref{eq:frequency_dependent_antenna_patterns_time_detector_response} is that the round-trip time $2T$ of the photon in each arm cavity is a parameter of the detector response\footnote{We wish to stress that the time--dependency described here is unrelated to the motion of the detector in space, such as the spacecraft orbital motion of the LISA constellation which is well described, e.g., in \cite{marsat2018fourier}.}. Moreover, the gravitational wave signal is not necessarily constant during the travel time of the photon, as it was in eq. \eqref{eq:detector_response_first_formula} (which can be recovered from eq. \eqref{eq:frequency_dependent_antenna_patterns_time_detector_response} assuming $h(t)$ to be constant over the photon round-trip $2T$), implying a more complex interplay between strain and detector geometry.

The total photon round-trip time is split between $T_{+}^{\hat{\textbf{a}}}$ and $T_{-}^{\hat{\textbf{a}}}$, and is a function of the sky location of the GW source with respect to the interferometer arms. This split introduces an additional dependence of the detector response on sky location, which adds to that of the polarization tensors $e^{A}_{ij}(\hat {\textbf{n}})$.

It is worth mentioning that eq. \eqref{eq:frequency_dependent_antenna_patterns_time_detector_response} is computed over a single round trip of the photons inside the arms. This may come as unexpected, because the Fabry-Perot cavity \cite{black2003introduction}, brings the average number of round trips of a single photon in the range 50 to 100, depending on the finesse of the detector \cite{maggiore2008gravitational}. We discuss this aspect further in Section B in the Supplementary Text.

The frequency-dependence of the antenna patters has the major consequence that in the time domain the polarizations can no longer be disentangled from the detector geometry. This can be well appreciated by comparing the detector response in the time domain with that in the frequency domain (equations \eqref{eq:frequency_dependent_antenna_patterns_frequency_detector_response}, \eqref{eq:frequency_dependent_antenna_pattern_plus_definition} and \eqref{eq:frequency_dependent_antenna_pattern_cross_definition}). 
We further elaborate on this in Section C in the Supplementary Text.

\subsection{Impact of the frequency-dependent antenna patterns on analyses in the time-frequency domain}

We start considering a time-frequency representation of GW signals $s(t)$ obtained with continuous wavelets, where the wavelet transform $T$ of a given signal evaluated at the time-frequency position $(\tau,\nu)$ is

\begin{subequations}
\begin{align}
    T(\tau,\nu)_{s}&=\int_{-\infty}^{+\infty} dt \, s(t)\psi^{*}(t;\tau,\nu) \\
    &=\int_{-\infty}^{+\infty} df \, \tilde{s}(f)\tilde{\psi}^{*}(f;\tau,\nu),
\end{align}
 \end{subequations}
where $\psi^{*}(t;\tau,\nu)$ is the wavelet \cite{daubechies1992ten}, and where the tilde denotes its Fourier transform.

The transform of the detector response $\xi_{F}$ in the case of frequency-independent antenna patterns can be split into two polarization-dependent terms as follows
\begin{subequations}
\begin{align}
    &T_{\xi_{F}}(\tau,\nu)=\int_{-\infty}^{+\infty}df \, \tilde{\xi}_{F}(f,\hat{\textbf{n}})\tilde{\psi}^{*}(f;\tau,\nu)\\
    &=\sum_{A=+,\times}F_{A}(\hat{\textbf{n}}) \int_{-\infty}^{+\infty}df \,  \tilde{h}_{A}(f)\tilde{\psi}^{*}(f;\tau,\nu)\\
    &=\sum_{A=+,\times}F_{A}(\hat{\textbf{n}})T_{h_{A}}(\tau,\nu).
\end{align}    
\end{subequations}
This is no longer true in the case of frequency-dependent antenna patterns:
\begin{subequations}\label{eq:frequency_dependent_antenna_patterns_detector_response_TF_frequency}
\begin{align}
    &T_{\xi_{G}}(\tau,\nu)=\int_{-\infty}^{+\infty}df \, \tilde{\xi}_{G}(f,\hat{\textbf{n}})\tilde{\psi}^{*}(f;\tau,\nu)\\
    &=\sum_{A=+,\times} \int_{-\infty}^{+\infty}df \, \tilde{G}_{A}(f,\hat{\textbf{n}})\tilde{h}_{A}(f) \tilde{\psi}^{*}(f;\tau,\nu).
\end{align}    
\end{subequations}
From equations \eqref{eq:frequency_dependent_antenna_patterns_detector_response_TF_frequency} we see that it is no longer possible to express the complete wavelet transform as a sum of weighted transforms of the polarization components, with weights equal to the antenna patterns. We find instead that the weights are part of the integrals, which turn out to be dependent on the sky location.

It can also be shown (see Section D of the Supplementary Text) that starting from equations \eqref{eq:frequency_dependent_antenna_patterns_detector_response_TF_frequency} we obtain the following expression, which is nearly the same as the time-domain representation
\begin{widetext}
\begin{align}\label{eq:frequency_dependent_antenna_patterns_time_frequency_detector_response_time}
    T_{\xi_{G}}(\tau,\nu)=\sum_{A=+,\times}&\dfrac{1}{2}\epsilon_{11}^{A}(\hat{\textbf{n}}) 
    \left(  \dfrac{1}{2T_{-}^{\hat{\textbf{x}}}}\int_{\tau-2T}^{\tau-T_{+}^{\hat{\textbf{x}}}} T_{h_{A}}(t',\nu)d t'
    +\dfrac{1}{2T_{+}^{\hat{\textbf{x}}}}\int_{\tau-T_{+}^{\hat{\textbf{x}}}}^{\tau} T_{h_{A}}(t',\nu)d t' \right) \nonumber \\
    -&\dfrac{1}{2}\epsilon_{22}^{A}(\hat{\textbf{n}})  
    \left(  \dfrac{1}{2T_{-}^{\hat{\textbf{y}}}}\int_{\tau-2T}^{\tau-T_{+}^{\hat{\textbf{y}}}} T_{h_{A}}(t',\nu)d t'
    +\dfrac{1}{2T_{+}^{\hat{\textbf{y}}}}\int_{\tau-T_{+}^{\hat{\textbf{y}}}}^{\tau} T_{h_{A}}(t',\nu)d t' \right).
\end{align}    
\end{widetext}
This key result displays the deep link existing between the time-domain and the time-frequency domain representations. In neither cases it is possible to disentangle the polarized strains and the antenna patterns
This issue is extremely relevant for likelihood-based unmodeled analyses, where the ability to disentagle the polarizations is crucial, as we discuss in the following section.

\section{Impact of the frequency-dependent antenna patterns on unmodeled analysis}

\subsection{Likelihood-based unmodeled analysis}\label{sec:unmodeled_analysis}

Unmodeled analysis exploits the coherence of GW signals in different detectors to find the polarised GW strains, and this can be done using a single likelihood function that combines all the detector responses \cite{klimenko2016method}. Assuming that experimental uncertainties are independent and are distributed according to a unit-variance Gaussian probability density function (pdf) after whitening, one finds that maximising the likelihood is equivalent to minimising the following chi-square in the frequency or time--frequency domain 
\begin{equation}
S = \sum_{k=1}^K \left| \tilde{w}_k -\left(  \tilde{g}_{k+} \tilde{h}_+ + \tilde{g}_{k\times} \tilde{h}_\times \right) \right|^2,
\label{eq:least-squares}
\end{equation}
where $K$ denotes the number of detectors, labelled as $k=1,...,K$, $w_k$ is the whitened response of the $k$-th detector with Fourier components $\tilde{w}_k = \tilde{\xi}_k(f)/\sigma_k(f)$, $\sigma_k(f)$ is the power spectral density (PSD) of the $k$-th detector, and the $\tilde{g}_{k,A}$s are the whitened frequency-dependent antenna patterns $\tilde{g}_{k,A}(f) = G_{k,A}(f)/\sigma_k(f)$.

In the frequency domain it is convenient to write these quantities in vector form, so that the whitened detector response reads
\begin{equation}\label{eq:whitened_detector_response_modified_antenna_patterns}
    \tilde{\boldsymbol{w}}(f,\hat{\textbf{n}})=\tilde{\boldsymbol{g}}_{+}(f,\hat{\textbf{n}})\tilde{h}_{+}(f)+\tilde{\boldsymbol{g}}_{\times}(f,\hat{\textbf{n}})\tilde{h}_{\times}(f)+\tilde{\boldsymbol{\zeta}}(f).
\end{equation}
where $\tilde{\boldsymbol{\zeta}}(f)$ is the whitened noise array in the frequency domain. 

\medskip

We find the whitened, polarised strains $h_{+}$ and $h_{\times}$ by minimising the chi-square \eqref{eq:least-squares} with respect to the two polarizations $h_{+}$ and $h_{\times}$.

Most unmodeled pipelines solve the LWA version of eq. \eqref{eq:least-squares} in the time-frequency domain \cite{klimenko2016method, cornish2021bayeswave}, but this is no longer possible with frequency--dependent antenna patterns because in the time-frequency domain the polarized strains cannot be disentagled from the antenna patterns, as discussed in the previous Section.
Therefore, the minimization of eq. \eqref{eq:least-squares} with respect to $\tilde{h}_{+}$ and $\tilde{h}_{\times}$ can only be performed in the frequency domain.  Straightforward algebraic manipulations lead us to the frequency-domain solutions 
\begin{subequations}
\label{eq:fdsolutions}
\begin{align}    \tilde{h}^{r}_{+}&=\dfrac{\tilde{\boldsymbol{g}}_{+}^{\dagger}|\tilde{\boldsymbol{g}}_{\times}|^{2}-\tilde{\boldsymbol{g}}_{\times}^{\dagger}(\tilde{\boldsymbol{g}}^{\dagger}_{+}\tilde{\boldsymbol{g}}_{\times})}{|\tilde{\boldsymbol{g}}_{+}|^{2}|\tilde{\boldsymbol{g}}_{\times}|^{2}-|\tilde{\boldsymbol{g}}^{\dagger}_{+}\cdot\tilde{\boldsymbol{g}}_{\times}|^{2}} \cdot\tilde{\boldsymbol{w}}
\label{eq:plus_polarization_solution_modified_antenna_patterns_generic}\\   \tilde{h}^{r}_{\times}&=\dfrac{\tilde{\boldsymbol{g}}_{\times}^{\dagger}|\tilde{\boldsymbol{g}}_{+}|^{2}-\tilde{\boldsymbol{g}}_{+}^{\dagger}(\tilde{\boldsymbol{g}}^{\dagger}_{\times}\tilde{\boldsymbol{g}}_{+})}{|\tilde{\boldsymbol{g}}_{+}|^{2}|\tilde{\boldsymbol{g}}_{\times}|^{2}-|\tilde{\boldsymbol{g}}^{\dagger}_{+}\cdot\tilde{\boldsymbol{g}}_{\times}|^{2}}\cdot\tilde{\boldsymbol{w}},
\label{eq:cross_polarization_solution_modified_antenna_patterns_generic}  
\end{align}
\end{subequations}
where for readability we have dropped the explicit dependence on $f$ and $\hat{\textbf{n}}$, and where the superscript $r$ indicates that these are the reconstructed polarizations.
Beyond the trivial fact that compared to the solution found for frequency-independent antenna patterns \cite{di2018estimation} all quantities are complex numbers (among them, the antenna patterns), there is also another difference worth noticing.
Many pipelines, such as cWB \cite{klimenko2016method, drago2021coherent} and the X-Pipeline \cite{sutton2010x}, carry out their calculation within a specific polarization frame, the Dominant Polarization Frame (DPF), for which a specific polarization angle $\psi$ is chosen such that the sensitivity to the specific $+$ polarization is maximized.
As a result of this choice, the antenna pattern arrays to became orthogonal, i.e., with the usual frequency--independent antenna patterns, $\tilde{\boldsymbol{f}}_+(\psi) \cdot \tilde{\boldsymbol{f}}_\times(\psi) = 0$, leading to a simplified form of eqs. \eqref{eq:fdsolutions}; moreover, it provides an elegant geometric visualization of the unmodeled formalism, which is well discussed, e.g., in \cite{drago2010search}. 
The same simplification cannot be achieved with frequency--dependent antenna patterns, due to the fact that these are complex numbers: indeed, the maximization of the sensitivity to the $+$ polarization implies the orthogonality of the antenna patterns as usually understood in the complex domain, i.e., $\text{Re}\left[ \tilde{\boldsymbol{g}}^{\dagger}_+(\psi) \cdot \tilde{\boldsymbol{g}}_\times(\psi)\right] = 0$,  which no longer leads to a simplified form of eqs. \eqref{eq:fdsolutions}.
Moreover, because of the frequency dependence of the antenna patterns the DPF also becomes intrinsically frequency dependent, which means that different frequencies correspond to different polarization angles.
Additional details on this topic can be found in the Supplementary Text, Section E.

\medskip

To complete our unmodeled formalism, using eqs. \eqref{eq:fdsolutions} we note that in the frequency domain the reconstructed responses, i.e. the quantities 
\begin{equation}
\label{eq:wrscalar}
\tilde{s}_k^{(r)} = \tilde{g}_{+k} \tilde{h}_+^{(r)} + \tilde{g}_{\times k} \tilde{h}_\times^{(r)}
\end{equation}
can be set in the very compact form 
\begin{equation}\label{eq:modified_antenna_patterns_reconstructed_detector_response}
    \tilde{\boldsymbol{s}}^{r}=\mathcal{G}\tilde{\boldsymbol{w}}
\end{equation}
where the matrix ${\mathcal{G}}$ is
\begin{widetext}
\begin{equation}\label{modified_antenna_patterns_matrix}
    \mathcal{G}_{kl}=\dfrac{\left( |\tilde{\boldsymbol{g}}_{\times}|^{2}\tilde{g}_{l+}^{*}-(\tilde{\boldsymbol{g}}^{\dagger}_{+}\cdot\tilde{\boldsymbol{g}}_{\times})\tilde{g}_{l\times}^{*} \right)\tilde{g}_{k+} + \left( |\tilde{\boldsymbol{g}}_{+}|^{2}\tilde{g}_{l\times}^{*}-(\tilde{\boldsymbol{g}}^{\dagger}_{\times}\cdot\tilde{\boldsymbol{g}}_{+})\tilde{g}_{l+}^{*} \right)\tilde{g}_{k\times}}{|\tilde{\boldsymbol{g}}{+}|^{2}|\tilde{\boldsymbol{g}}_{\times}|^{2}-|\tilde{\boldsymbol{g}}^{\dagger}_{+}\tilde{\boldsymbol{g}}_{\times}|^{2}}.
\end{equation}    
\end{widetext}
In the next sections we discuss the dangers of ignoring the frequency-dependent antenna patterns, and show how to carry out a correct analysis and introduce revised definitions of coherent energy and noise energy \cite{sutton2010x}.

\subsection{Biases due the use of frequency-independent antenna patterns in the unmodeled analysis of next--generation detectors' data}\label{sec:waveform_reconstruction_biases}

\onecolumngrid

\begin{figure}[!ht]
    \centering
    \includegraphics[width=1.\textwidth]{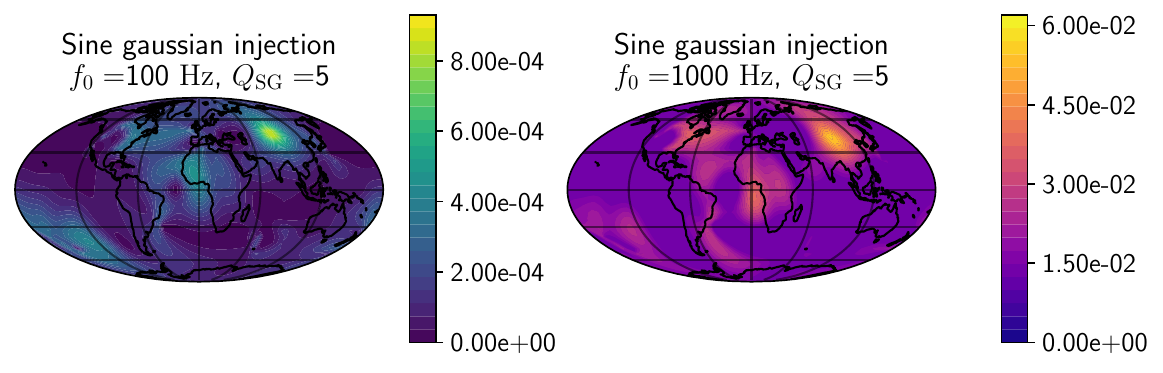}
    \caption{Mismatch between injected and reconstructed waveform, evaluated at different source locations, for the detector network ET1\_L, ET2\_L, CE1 and CE2, with reconstruction performed involving frequency--independent antenna patterns.
    The injected signals are SGs with $Q_{SG}=5$ and $f_{0}=\SI{100}{Hz},\,\SI{1000}{Hz}$ at different source locations.}  \label{fig:mismatch_Q_5_CE1_CE2_ET1_L_ET2_L_detectors}
\end{figure}{}  

\begin{figure}[ht]
    \centering
    \includegraphics[width=1.\textwidth]{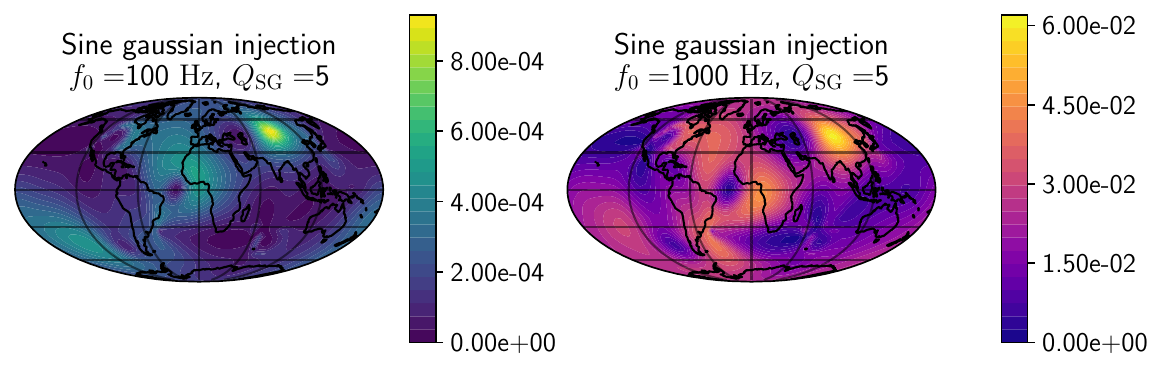}
    \caption{Mismatch between injected and reconstructed waveform, evaluated at different source locations, for the detector network ET1\_tr, ET2\_tr, ET3\_tr, CE1 and CE2, with reconstruction performed involving frequency--independent antenna patterns.
    The injected signals are SGs with $Q_{SG}=5$ and $f_{0}=\SI{100}{Hz},\,\SI{1000}{Hz}$ at different source locations.}  \label{fig:mismatch_Q_5_CE1_CE2_ET1_tr_ET2_tr_ET3_tr_detectors}
\end{figure}{}

\twocolumngrid

\begin{figure}[!ht]
    \centering
    \includegraphics[width=.47\textwidth]{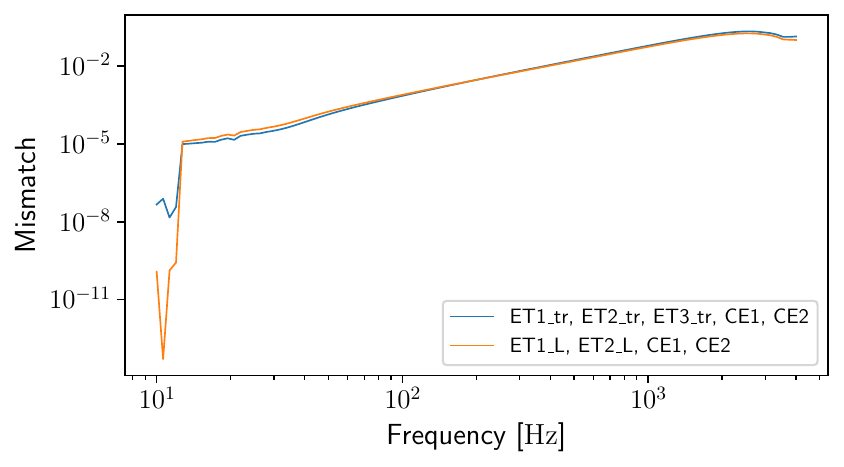}
    \caption{Mismatch between injected and reconstructed waveform, evaluated as a function of the SG central frequency, for both the detector networks ET1\_L, ET2\_L, CE1, CE2 and ET1\_tr, ET2\_tr, ET3\_tr, CE1, CE2, with reconstruction performed involving frequency-independent antenna patterns.
    For each detector network, the source location of the injected signal is that giving the highest mismatch.
    The $Q$ parameter of the SG signal is fixed at $Q_{SG}=5$.}  \label{fig:mismatch_vs_frequency}
\end{figure}{}  

\begin{figure}[!ht]
    \centering
    \includegraphics[width=.47\textwidth]{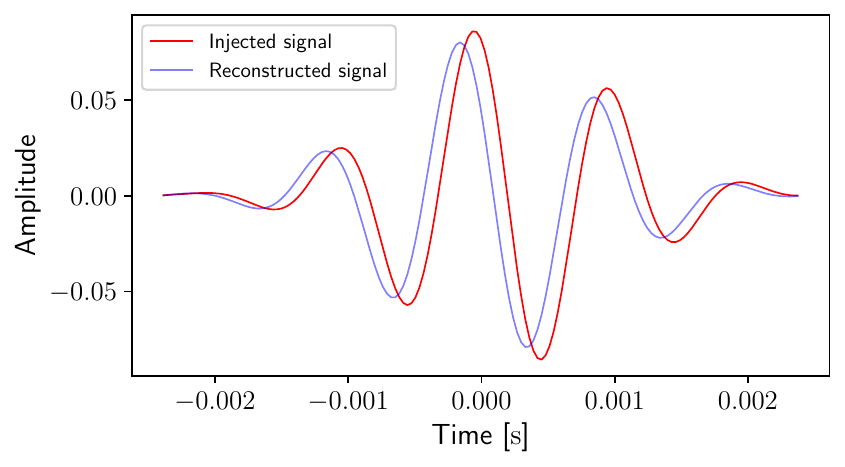}
    \caption{An example of injected and reconstructed waveform obtained injecting for the detector network ET1\_L, ET2\_L, CE1 and CE2 a SG signal with $Q_{SG}=5$, $f_{0}=\SI{1000}{Hz}$ at the source location giving the highest mismatch for that detector network, and then reconstructing it with frequency-independent antenna patterns. The represented waveforms are those of the CE1 detector.}  \label{fig:mismatch_waveform}
\end{figure}{} 

The formalism associated with the DPF and concepts like coherent energy and noise energy are well-established in the world of unmodeled analysis and we could be tempted to stick to them neglecting what may appear to be just a weak frequency dependence of the antenna patterns. In this Section we examine the consequences of such a choice by evaluating its impact on waveform reconstruction. We also verify that by using frequency--dependent antenna patterns and the corresponding unmodeled formalism, the reconstructions are not biased. 

\medskip 

To this end, we carry out the following steps:
\begin{enumerate}
    \item Using the expression of the time-domain detector response, eq. \eqref{eq:frequency_dependent_antenna_patterns_time_detector_response}, we compute the response of the detectors to a sine-Gaussian (SG) signal defined by the polarized strains 
    \begin{subequations}
    \begin{align}
     h_{+}(t)&=A e^{-(t /\tau)^{2}}\sin{(2\pi f_{0}t)} \\
     h_{\times}(t)&=A e^{-(t/\tau)^{2}}\cos{(2\pi f_{0}t)}
    \end{align}
    \end{subequations}
    where $A$ is the amplitude of the signal, $\tau=Q_{SG}/(\sqrt{2}\pi f_{0})$ is its characteristic time, and $f_{0}$ its central frequency \cite{drago2010search,abadie2012all}. Throughout these tests we take $A=10^{-23}$ and $Q_{SG}=5$ (this is not a restrictive condition as we have verified that the dependence on $Q_{SG}$ is very weak and that the results appear nearly identical in the range from $Q_{SG}=5$ and $Q_{SG}=100$).
    We also assume a well-defined sky location of the GW source $\hat{\textbf{n}}$ and that it is detected by two different detector networks: the first, includes the CE detectors and the ET 2L design (CE1, CE2, ET1\_L and ET2\_L); the second, includes the CE detectors and the ET triangular design (CE1, CE2, ET1\_tr, ET2\_tr and ET3\_tr). 
    We remark here that all injections performed in this paper are only meant to map the impact of frequency dependence on different sky locations and do not include noise; however, we do include signal whitening\footnote{For a description of whitening as performed by several pipelines of the LVK Collaboration, see the GWpy documentation \cite{gwpy,GWpyWhiten}.} because it modulates the frequency dependence.
    The detector sensitivities needed for data whitening are described in \cite{abbott2017exploring} (the numerical values are also available, see \cite{LIGODocumentP1600143}). Here, a single CE sensitivity has been assumed for all CE detectors, as well as a single ET sensitivity for all ET detectors.
    \item We reconstruct the detector responses, defined by eq. \eqref{eq:wrscalar}, using eq. \eqref{eq:modified_antenna_patterns_reconstructed_detector_response}, with both frequency--independent antenna patterns and with  frequency--dependent antenna patterns
    \item Next, we evaluate the biases in waveform reconstruction using the mismatch between the injected detector response $\boldsymbol{s}^{inj}$ and the reconstructed detector response $\boldsymbol{s}^{r}$, defined as one minus the overlap $\mathcal{O}$ between them \cite{salemi2019wider}
\begin{equation}\label{eq:mismatch}
    1-\mathcal{O}=1-\dfrac{(\boldsymbol{s}^{rec}, \boldsymbol{s}^{inj})}{\sqrt{(\boldsymbol{s}^{rec}, \boldsymbol{s}^{rec})} \cdot\sqrt{(\boldsymbol{s}^{inj}, \boldsymbol{s}^{inj})}},
\end{equation}
and where the scalar product between whitened waveforms is defined by $(\boldsymbol{a},\boldsymbol{b})=\sum_{k}\int a_{k}(t)b_{k}(t)dt$.
\end{enumerate}

To avoid distorsions due to an insufficient sampling rate\footnote{This is important because SG signals are not band-limited, although the low-frequency and high-frequency tails of their power spectral density decay very quickly.}, we fix it at four times the nearest power of $2$ higher than $f_{0}$: with $Q_{SG}=5$, this leads to a Nyquist frequency $\sim f_{0}+7\sigma_{f_{0}}$, with $\sigma_{f_{0}}=f_{0}/\sqrt{2}Q_{SG}$ the standard deviation of the Gaussian envelope in frequency domain. 
Finally, by looping over a grid of sky locations, we produce maps of the mismatch similar to the maps of antenna pattern difference shown in Figures 
\ref{fig:percentage_difference_100_Hz_CE_detectors}-\ref{fig:percentage_difference_1000_Hz_ET_detectors}.
Figures \ref{fig:mismatch_Q_5_CE1_CE2_ET1_L_ET2_L_detectors} and \ref{fig:mismatch_Q_5_CE1_CE2_ET1_tr_ET2_tr_ET3_tr_detectors} show the mismatch obtained reconstructing the SG with frequency-independent antenna patterns as a function of the source location for the network consisting of CE1, CE2, ET1\_L and ET2\_L detectors, and for the network consisting of CE1, CE2, ET1\_tr, ET2\_tr and ET3\_tr, respectively.
While the mismatch is negligible at $f=\SI{100}{Hz}$, this is not the case for $f=\SI{1000}{Hz}$, where many source locations produce a mismatch between $1\%$ and $6\%$. 
Even if not huge, this systematic mismatch should be no larger than that produced by noise, which may not be the case with the increased sensitivity of the next-generation detectors. 

Taking for each network of detectors the source location maximizing the mismatch, in Figure \ref{fig:mismatch_vs_frequency} we plot  this mismatch as a function of the central frequency of the SG signal.
Above $\sim \SI{1000}{Hz}$ the mismatch reaches values around $10\%$, which corresponds to a considerable bias in waveform reconstruction. 

We can see the effects of the mismatch on the waveform in Figure \ref{fig:mismatch_waveform}, where we compare the injected waveform  with the reconstruction obtained with frequency--independent antenna patterns.
Although the injected and reconstructed waveforms appear shifted in time relative to each other, it is worth noting that the shift between them is not necessarily the same at all sampling times, as can be seen from eq. \eqref{eq:frequency_dependent_antenna_patterns_time_detector_response}. Using the mean value theorem\footnote{$\frac{1}{b-a}\int_{a}^{b}f(x)dx=f(c\in [a,b])$}, eq. \eqref{eq:frequency_dependent_antenna_patterns_time_detector_response} can be written as
\begin{widetext}   
\begin{align}\label{eq:frequency_dependent_antenna_patterns_time_detector_response_mean_value}
    \xi_{G}(t,\hat{\textbf{n}})=\sum_{A=+,\times}\dfrac{1}{2}\epsilon_{11}^{A}(\hat{\textbf{n}}) 
    \dfrac{1}{2}&\left( h_{A}(t_{A}\in[t-2T,t-T_{+}^{\hat{\textbf{x}}}]) + h_{A}(t_{A}\in[t-T_{+}^{\hat{\textbf{x}}},t])\right)\nonumber \\ 
    -\dfrac{1}{2}\epsilon_{22}^{A}(\hat{\textbf{n}})\dfrac{1}{2}&\left( h_{A}(t_{A}\in[t-2T,t-T_{+}^{\hat{\textbf{y}}}]) + h_{A}(t_{A}\in[t-T_{+}^{\hat{\textbf{y}}},t])\right).
\end{align}
\end{widetext}
Therefore, the integrals of eq. \eqref{eq:frequency_dependent_antenna_patterns_time_detector_response} effectively shift the sampling time. In general, these shifts are different from each other so that at every $t$ the four polarized strains in the equation are sampled at different times.
The effect of this sampling shift is negligible in the LWA limit, but can be considerable at frequencies close to the free spectral range, causing possible biases, e.g. in source localization, as already investigated in \cite{essick2017frequency}.
Indeed, the time shifts produced by the frequency-dependent antenna patterns (which depends on arm length, frequency, and source location) add to the different arrival times at each detector, playing a role in source localization.

\subsection{Use of frequency-dependent antenna patterns in the unmodeled analysis of next-generation detectors' data}\label{sec:waveform_reconstruction_with_frequency_dependent_antenna_patterns}

\begin{figure}[!ht]
    \centering
    \includegraphics[width=1.\textwidth]{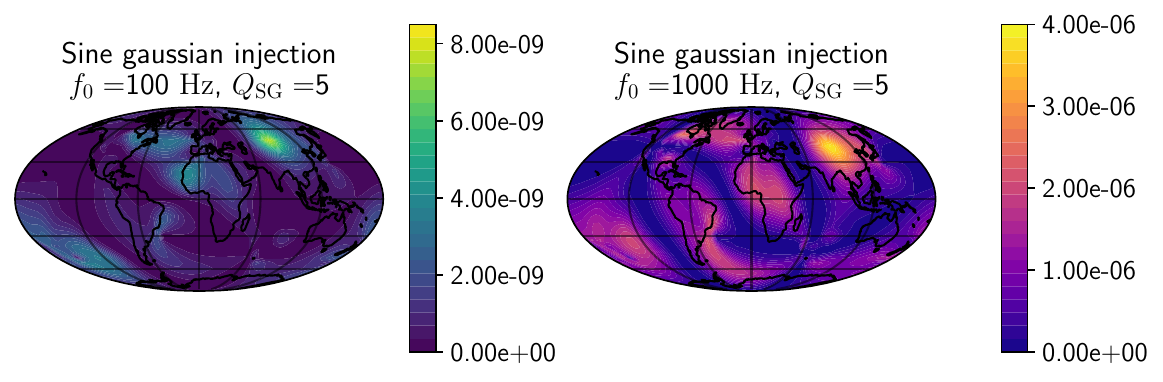}
    \caption{Mismatch between injected and reconstructed waveform, evaluated at different source locations, for the detector network ET1\_L, ET2\_L, CE1 and CE2, with reconstruction performed involving frequency-dependent antenna patterns.
    The injected signals are SGs with  $Q_{SG}=5$ and $f_{0}=\SI{100}{Hz},\,\SI{1000}{Hz}$ at different source locations.}  \label{fig:map_mismatch_Q_5_CE1_CE2_ET1_L_ET2_L_detectors}
\end{figure}

\begin{figure}[!ht]
    \centering
    \includegraphics[width=1.\textwidth]{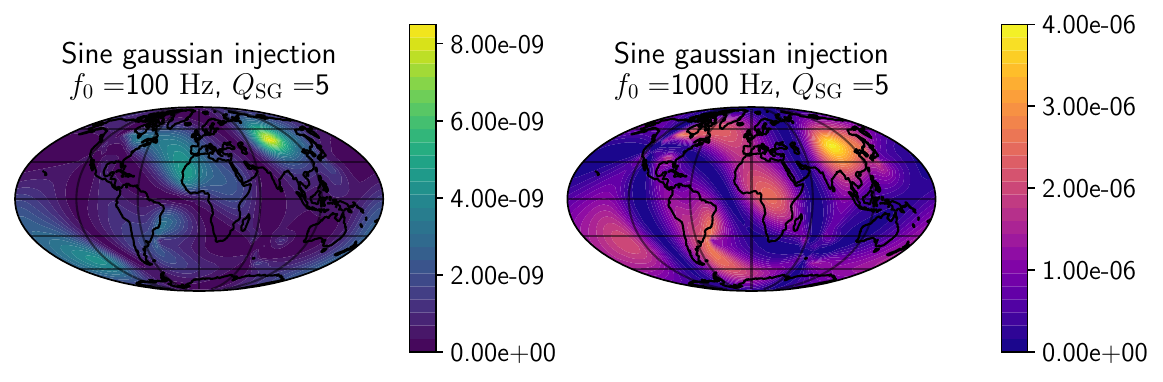}
    \caption{Mismatch between injected and reconstructed waveform, evaluated at different source locations, for the detector network ET1\_tr, ET2\_tr, ET3\_tr, CE1 and CE2, with reconstruction performed involving frequency-independent antenna patterns.
    The injected signals are SGs with  $Q_{SG}=5$ and $f_{0}=\SI{100}{Hz},\,\SI{1000}{Hz}$ at different source locations.}  \label{fig:map_mismatch_Q_5_CE1_CE2_ET1_tr_ET2_tr_ET3_tr_detectors}
\end{figure}

\begin{figure}[]
    \centering
    \includegraphics[width=.47\textwidth]{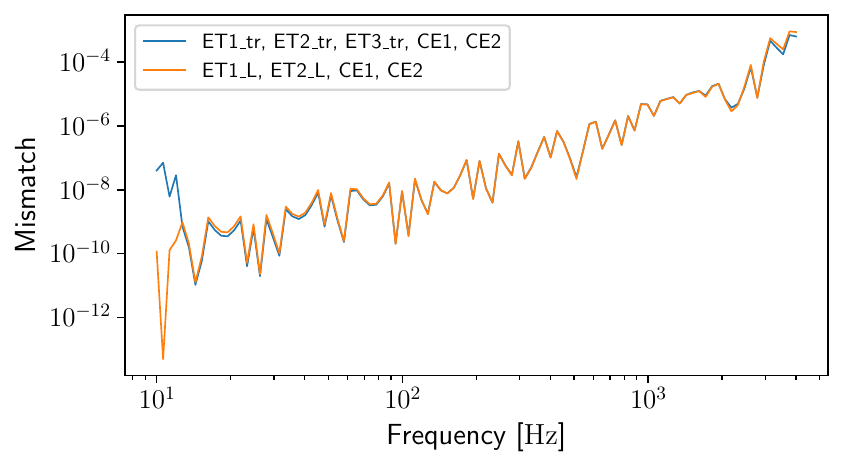}
    \caption{Mismatch between injected and reconstructed waveform, evaluated as a function of SG central frequency, for both the detector networks ET1\_L, ET2\_L, CE1, CE2 and ET1\_tr, ET2\_tr, ET3\_tr, CE1, CE2, with reconstruction performed involving frequency-dependent antenna patterns.
    For each detector network, the source location of the injected signal is that giving the highest mismatch.
    The $Q$ parameter of the SG signal is fixed at $Q_{SG}=5$.}  \label{fig:map_mismatch_vs_frequency}
\end{figure}

\begin{figure}[]
    \centering
    \includegraphics[width=.47\textwidth]{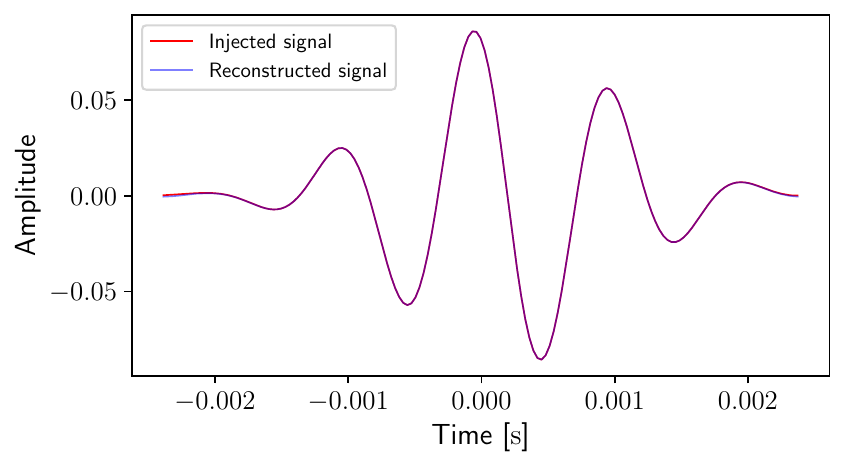}
    \caption{An example of injected and reconstructed waveform obtained injecting for the detector network ET1\_L, ET2\_L, CE1 and CE2 a SG signal with $Q_{SG}=5$, $f_{0}=\SI{1000}{Hz}$ at the source location giving the highest mismatch for that detector network, and then reconstructing it with frequency-dependent antenna patterns. The represented waveforms are those of the CE1 detector.}  \label{fig:map_mismatch_waveform}
\end{figure}

In this short Section we show that with frequency--dependent antenna patterns the reconstructed waveforms display no relevant biases.
This is apparent in Figures \ref{fig:map_mismatch_Q_5_CE1_CE2_ET1_L_ET2_L_detectors} and \ref{fig:map_mismatch_Q_5_CE1_CE2_ET1_tr_ET2_tr_ET3_tr_detectors} which show the mismatch as a function of source location, while Figure \ref{fig:map_mismatch_vs_frequency} shows the mismatch as a function of the central frequency of the SG signal and Figure \ref{fig:map_mismatch_waveform} which shows an injected waveform and the corresponding reconstructed waveform.
The resulting mismatch is negligible compared to that obtained with frequency--independent antenna patterns and is likely the result of numerical roundoff errors.
Therefore, even if the polarizations and the antenna patterns are no longer disentangled in the time domain, their disentanglement in the frequency domain allows the signal waveform to be reconstructed by unmodeled methods with minor adjustments to the corresponding formalism.

In summary, the LWA can be used in next--generation detectors only at sufficiently low frequencies: the analysis of signals above  $\sim \SI{1000}{Hz}$ with frequency--independent antenna patterns could lead to significant distortions in  waveform reconstruction, depending also on the source location in the sky.

\onecolumngrid

{}

{}

{}  

{}

\begin{figure}[!ht]
    \centering
    \includegraphics[width=0.9\textwidth]{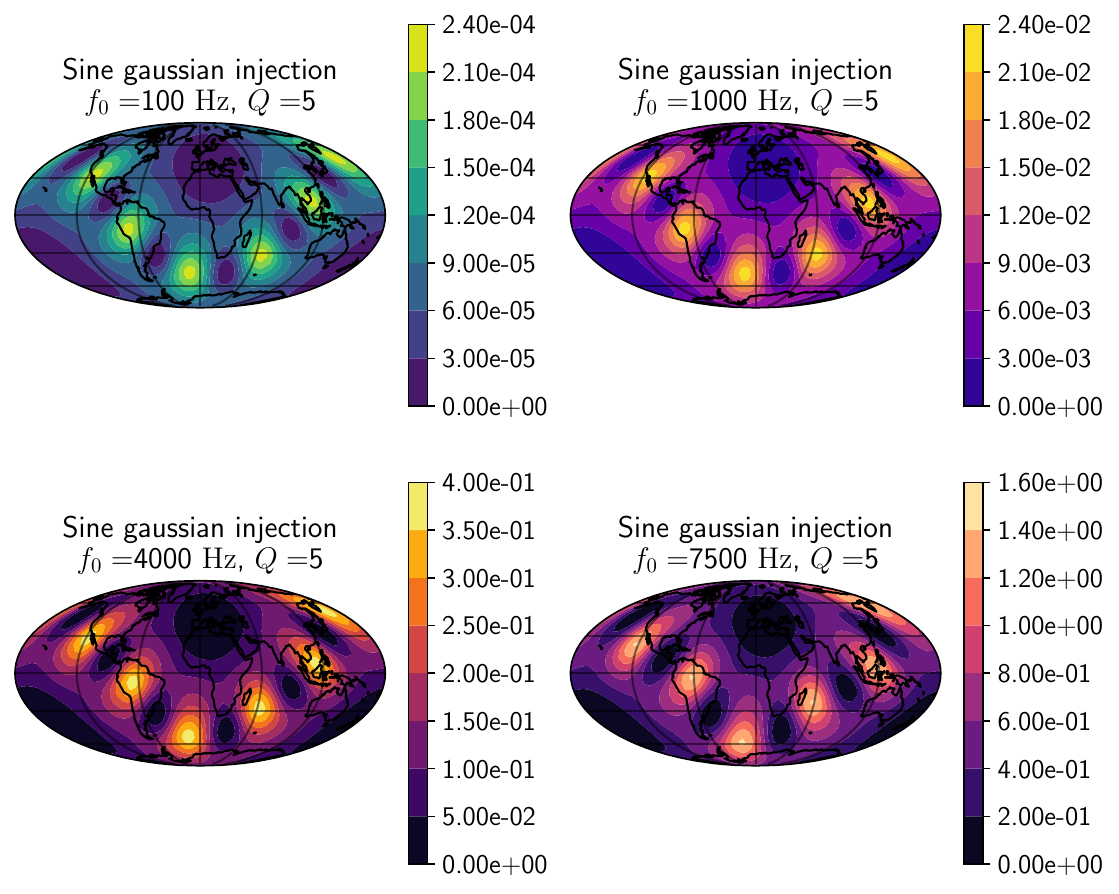}
    \caption{For the ET triangular design, null stream energy over the signal energy in the detectors (as in eq. \eqref{eq:ET_tr_null_stream}), for SG injections with $Q_{SG}=5$ and $f_{0}=\SI{100}{},\SI{1000}{},\SI{4000}{},\SI{7500}{Hz}$, as a function of source location.}  \label{fig:null_stream_Q_5_ET_triangular_detectors}
\end{figure}{}  

\twocolumngrid

\begin{figure}[!ht]
    \centering
    \includegraphics[width=0.47\textwidth]{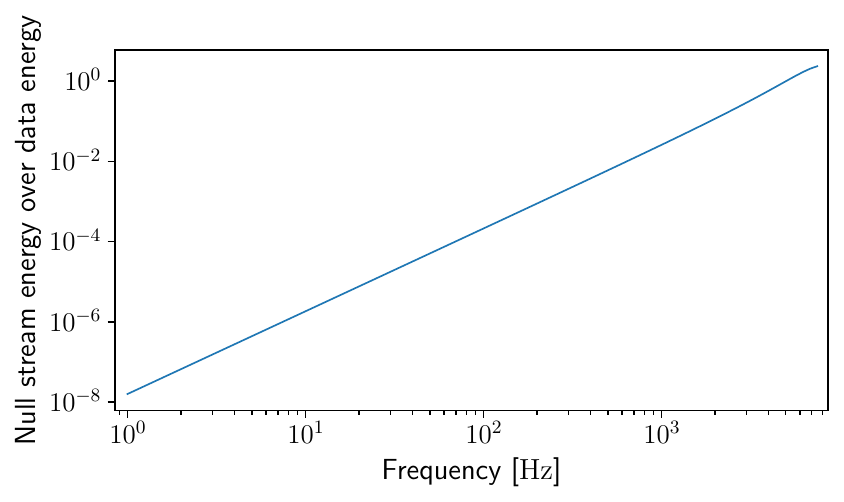}
    \caption{For the ET triangular design, null stream energy over the signal energy in the detectors (as in eq. \eqref{eq:ET_tr_null_stream}), for SG injections with $Q_{SG}=5$ and fixed source location, as a function of the  central frequency of the SG signal.}  \label{fig:null_stream_Q_5_ET_triangular_detectors_vs_frequency}
\end{figure}{}

\section{Impact on the null stream of ET (triangular design)}\label{sec:ET_null_stream}

The aim of this Section is to understand the impact of the frequency--dependent antenna patterns on the source--location independent null stream planned for the ET triangular design.
The usefulness of this concept depends on the fact that in the low frequency limit --- therefore, with frequency--independent antenna patterns -- the sum of the responses of the three interferometers in the triangular design cancels out the signal, i.e.,
\begin{small}
\begin{align}\label{eq:usual_antenna_patterns_ET_null_stream}
&\sum_{k=1,3}  \xi_{F}^{ET\_tr,k} =\nonumber \\ 
&=\left[ \sum_{k=1,3}  F_{+}^{ET\_tr,k } \right] h_{+} + \left[ \sum_{k=1,3}  F_{\times}^{ET\_tr,k }\right] h_{\times}=0.       
\end{align}
\end{small}
This topic has been investigated multiple times (see, e.g., \cite{regimbau2012mock} which describes this null stream as  ``an invaluable tool in data analysis", and \cite{wong2022signal,goncharov2022utilizing,janssens2022formulation}): indeed, the source--location independent null stream allows characterizing the combined noise properties simply by summing the responses of the three interferometers, and this can be useful for many tasks such as power spectral density estimation.
Some doubts still linger about the practical implementation of this null stream, mainly due to the noise correlation of nearby interferometers, such as those in the triangular design \cite{branchesi2023science}. Here, we  point to yet another problem due to the limited applicability of the LWA to signals detected by ET. 

First, we recall the frequency--dependent antenna patterns for the ET triangular design (see Section F in the Supplementary Text)
\begin{widetext}
\begin{equation}
    \tilde{G}_{A}^{\ang{60}}(f,\hat{\textbf{n}})=
    \dfrac{1}{2}\left[\epsilon_{11}^{A}(\hat{\textbf{n}})\tilde{D}(\hat{\textbf{x}}, \hat{\textbf{n}}, f) - \left( \dfrac{1}{4} \epsilon_{11}^{A}(\hat{\textbf{n}}) +\dfrac{3}{4} \epsilon_{22}^{A}(\hat{\textbf{n}}) +\dfrac{\sqrt{3}}{4} \left( \epsilon_{12}^{A}(\hat{\textbf{n}})+\epsilon_{21}^{A}(\hat{\textbf{n}}) \right)\right)\tilde{D}(\hat{\textbf{y}}, \hat{\textbf{n}}, f)  \right]\label{eq:ET_modified_antenna_patterns_general_formula}
\end{equation}   
where, just as before, the frequency-domain response of each $\ang{60}$ interferometer is $\xi_{G}^{\ang{60}}(f,\hat{\textbf{n}})=\sum_{A=+,\times} G_{A}(f,\hat{\textbf{n}}) h_{A}(f)$.
The corresponding time-domain response is
\begin{equation}\label{eq:ET_modified_antenna_patterns_time_domain_general_formula_first_version}
    \xi_{G}^{\ang{60}}(t,\hat{\textbf{n}})=\dfrac{1}{2}\sum_{A=+,\times}\left[\epsilon_{11}^{A}(\hat{\textbf{n}})\Bar{h}_{A}(t,\hat{\textbf{n}},\hat{\textbf{x}}) - \left( \dfrac{1}{4} \epsilon_{11}^{A}(\hat{\textbf{n}}) +\dfrac{3}{4} \epsilon_{22}^{A}(\hat{\textbf{n}}) +\dfrac{\sqrt{3}}{4} \left( \epsilon_{12}^{A}(\hat{\textbf{n}})+\epsilon_{21}^{A}(\hat{\textbf{n}}) \right)\right)\Bar{h}_{A}(t,\hat{\textbf{n}},\hat{\textbf{y}})  \right]
\end{equation}    
\end{widetext}
with 
\begin{equation}
    \Bar{h}_{A}(t,\hat{\textbf{n}},\hat{\textbf{a}})=\dfrac{1}{2T_{-}^{\hat{\textbf{a}}}}\int_{t-2T}^{t-T_{+}^{\hat{\textbf{a}}}} h_{A}(\tau)d\tau
    +\dfrac{1}{2T_{+}^{\hat{\textbf{a}}}}\int_{t-T_{+}^{\hat{\textbf{a}}}}^{t} h_{A}(\tau)d\tau.\label{eq:polarization_integrals}
\end{equation}
Eq. \eqref{eq:usual_antenna_patterns_ET_null_stream} is no longer true when we replace the frequency--independent with frequency--independent antenna patterns. 
\begin{equation}\label{eq:modified_antenna_patterns_ET_null_stream}   \xi^{ET1\_tr}_{G}+\xi^{ET2\_tr}_{G}+\xi^{ET3\_tr}_{G}\neq 0 .
\end{equation}
Clearly, the departure from condition \eqref{eq:usual_antenna_patterns_ET_null_stream} grows gradually with frequency, and here we evaluate the null-stream energy as a function of frequency by means of repeated injection of SG signals. We consider the following ratio of signal energies
\begin{equation}
\epsilon_{r}=\dfrac{\int\left[\xi^{ET1\_tr}_{G}(t)+\xi^{ET2\_tr}_{G}(t)+\xi^{ET3\_tr}_{G}(t)\right]^{2}\, dt}{\int\left[\xi^{ET1\_tr}_{G}(t)^{2}+\xi^{ET2\_tr}_{G}(t)^{2}+\xi^{ET3\_tr}_{G}(t)^{2}\right]\, dt},\label{eq:ET_tr_null_stream}
\end{equation}
i.e. the overall null stream energy over the signal energy in the detectors.

The procedure follows a scheme similar to that already used in Section \ref{sec:waveform_reconstruction_biases}: first, we inject SG signals varying mainly in source location, considering only a few central frequencies ($f_{0}=\SI{100}{},\SI{1000}{},\SI{4000}{},\SI{7500}{Hz}$, to cover the whole frequency range planned fo the ET triangular design \cite{rowlinson2021feasibility}); then, we fix the source location by maximizing eq. \eqref{eq:ET_tr_null_stream} for $f_{0}=\SI{7500}{Hz}$. 
As before, for all the injections we consider $Q_{SG}=5$, we do not assume noise and we do include the whitening of signals.

The results are reported in Figures \ref{fig:null_stream_Q_5_ET_triangular_detectors} and \ref{fig:null_stream_Q_5_ET_triangular_detectors_vs_frequency}: it is interesting to note that also in this case, as shown several times in this paper, there is a significant dependence on the source location. 
In general, while in the LWA range the null stream is practically zero, just as expected (see the upper left panel in Figure \ref{fig:null_stream_Q_5_ET_triangular_detectors}), this is no longer true above $\sim \SI{1000}{Hz}$, where the null stream energy can be more than $150\%$ larger than the signal energy for $f_{0}=\SI{7500}{Hz}$.
At such frequencies, the null stream would then receive considerable contributions from signals with a high SNR: that makes the null stream of the triangular design of ET quite difficult to use for tasks such as power spectral density estimation.

\section{Conclusions}\label{sec:conclusions}

Unmodeled analysis methods must use the coherence among different GW detectors to reduce wrong GW detections and artifacts to a minimum. In current pipelines, to maximize the efficiency of the method the coherence is not limited to the time domain but rather it is extended to the time--frequency domain. With enough GW detectors in the network and assuming the LWA, likelihood--based computational schemes in the time or time-frequency domain return the polarized components of the strain. In this paper we have discussed the difficulties that we are about to face as detectors become larger and more sensitive and the LWA must be abandoned. A short list includes: 
\begin{itemize}
\item a zero--order solution can be found assuming frequency--independent antenna patterns, which is however affected by systematic errors (this is a consequence of the relative smallness of the correction produced by the frequency--dependent antenna patterns)
\item for accuracy, the source localization must be performed with the full frequency--dependent antenna patterns
\item for accuracy, the waveform reconstruction must be performed with the full frequency--dependent antenna patterns
\item the DPF is no longer unique, as it becomes frequency--dependent
\item the disentangling of the polarized strains can only be done in the frequency domain (leading to an increased dependence on background noise)
\item the null stream is deeply impacted by the frequency--dependent antenna pattern, in particular the triangular design of ET loses the appeal of having a fixed null stream for coherent GW signals
\end{itemize}
All of this is of crucial importance as we move towards higher and higher GW detector sensitivity and get closer to the first detection of a burst--only event such as the explosion of a CCSN-Supernova.

\section*{Acknowledgements}

We would like to thank the LVK groups of Trieste, Trento, Roma and Zürich for useful discussions.
This research has made use of data or software obtained from the Gravitational Wave Open Science Center (\url{http://gwosc.org}), a service of LIGO Laboratory, the LIGO Scientific Collaboration, the Virgo Collaboration, and KAGRA. LIGO Laboratory and Advanced LIGO are funded by the United States National Science Foundation (NSF) as well as the Science and Technology Facilities Council (STFC) of the United Kingdom, the Max-Planck-Society (MPS), and the State of Niedersachsen/Germany for support of the construction of Advanced LIGO and construction and operation of the GEO600 detector. Additional support for Advanced LIGO was provided by the Australian Research Council. Virgo is funded, through the European Gravitational Observatory (EGO), by the French Centre National de Recherche Scientifique (CNRS), the Italian Istituto Nazionale di Fisica Nucleare (INFN) and the Dutch Nikhef, with contributions by institutions from Belgium, Germany, Greece, Hungary, Ireland, Japan, Monaco, Poland, Portugal, Spain. KAGRA is supported by Ministry of Education, Culture, Sports, Science and Technology (MEXT), Japan Society for the Promotion of Science (JSPS) in Japan; National Research Foundation (NRF) and Ministry of Science and ICT (MSIT) in Korea; Academia Sinica (AS) and National Science and Technology Council (NSTC) in Taiwan. EM acknowledges support by ICSC – Centro Nazionale di Ricerca in High Performance Computing, Big Data and Quantum Computing, funded by European Union – NextGenerationEU.

\bibliographystyle{unsrt}
\bibliography{bibliography.bib}

\end{document}